\documentclass[aps,twocolumn,prx,preprintnumbers,10pt]{revtex4-2}
\usepackage{xpatch}
\makeatletter
\patchcmd{\@ssect@ltx}
    {\addcontentsline{toc}{#1}{\protect\numberline{}#8}}
    {}
    {}
    {}
\makeatother
\pdfoutput=1
\usepackage{color}
\usepackage{enumitem}
\usepackage{mathtools}
\usepackage[dvipsnames]{xcolor}

\usepackage{hyperref}
\definecolor{amaranth}{rgb}{0.9, 0.17, 0.31}
\definecolor{forestForestGreen(web)}{rgb}{0.13, 0.55, 0.13}
\definecolor{blue(munsell)}{HTML}{005567}
\definecolor{oxfordblue}{rgb}{0.0, 0.2, 0.4}
\definecolor{bblue}{rgb}{0.0, 0.58, 0.71}
\hypersetup{
pdfstartview={FitH}, 
pdftitle={}, 
colorlinks=true, 
linkcolor=oxfordblue, 
citecolor=oxfordblue, 
filecolor=oxfordblue, 
urlcolor=oxfordblue
}
\usepackage{pgfplots}
\pgfplotsset{compat=1.18}
\usepackage{amssymb}
\usepackage{amsfonts}
\usepackage{graphicx}
\usepackage{epstopdf}
\usepackage{dcolumn}
\usepackage{amsmath}
\usepackage{latexsym,bm}
\usepackage{amsthm}
\usepackage{slashed}
\usepackage{float}
\usepackage{color}
\usepackage{url}
\usepackage{rotating}
\usepackage[normalem]{ulem}
\usepackage{faktor}
\usepackage{tikz-cd}
\usepackage{tensor}
\usepackage{array}

\newcommand{\paragraf}[1]{\par\smallskip\noindent{\bf #1}}

\usepackage{tabularx}
\usepackage{makecell}
\usepackage[normalem]{ulem}
\usepackage{changepage}
\numberwithin{equation}{section}

\usepackage{cancel}

\usepackage{makecell}
\newtheorem{theorem}{Theorem}

\usepackage{tikz}
\usepackage{diagbox}
\usetikzlibrary{positioning}
\usetikzlibrary{calc}
\usetikzlibrary{decorations.pathreplacing,calligraphy}

\usepackage{xstring}
\usetikzlibrary{decorations.pathmorphing} 
\usetikzlibrary{decorations.markings} 
\usetikzlibrary{arrows} 
\usetikzlibrary{shapes} 
\usetikzlibrary{matrix} 
\usetikzlibrary{positioning} 
\usepackage[english]{babel} 
\usepackage[autostyle]{csquotes}
\usepackage{pifont}
\usetikzlibrary{shapes.multipart}

\tikzset{->-/.style={decoration={
  markings,
  mark=at position .6 with {\arrow{Latex[length=1.5mm,width=1.5mm]}}},postaction={decorate}}}

\usepackage{quantikz}

\newcommand{\bea}{\begin{eqnarray}}
\newcommand{\eea}{\end{eqnarray}}
\newcommand{\be}{\begin{equation}}
\newcommand{\ee}{\end{equation}}
\newcommand{\ba}{\begin{aligned}}
\newcommand{\ea}{\end{aligned}}
\newcommand{\bit}{\begin{itemize}}
\newcommand{\eit}{\end{itemize}}
\newcommand{\ben}{\begin{enumerate}}
\newcommand{\een}{\end{enumerate}}

\newcommand{\id}{\text{id}}

\newcommand{\Z}{{\mathbb Z}}

\newcommand{\bC}{{\mathbb C}}

\newcommand{\cC}{\mathcal{C}}

\newcommand{\cR}{\mathcal{R}}
\newcommand{\cS}{\mathcal{S}}
\newcommand{\cT}{\mathcal{T}}

\definecolor{mmUltramarine}{HTML}{2B2EA8}
\definecolor{mmPine}{HTML}{0E5A3A}
\definecolor{mmEmerald}{HTML}{1B8A4E}
\definecolor{mmTeal}{HTML}{13716C}
\definecolor{mmOlive}{HTML}{5F7516}
\definecolor{mmTobacco}{HTML}{8A4F24}
\definecolor{mmCream}{HTML}{D6CDBD}
\newcommand{\colL}{mmPine}            
\newcommand{\colT}{mmOlive}           
\newcommand{\colR}{mmEmerald}         
\newcommand{\colB}{mmTeal}            
\newcommand{\colU}{oxfordblue}
\newcommand{\colUa}{oxfordblue}
\newcommand{\colAux}{mmTobacco}       
\newcommand{\colFace}{mmCream} 
\newcommand{\colUname}{light blue}          
\newcommand{\colUaname}{dark blue}         

\makeatother

\usepackage{physics}
\renewcommand{\ol}{\overline}

\newcommand{\bbI}{\mathbb I}

\newcommand{\diag}{\text{diag}}

\newcommand{\III}{\text{III}}
\usepackage{float}

\usepackage{makecell}

\makeatletter
\newcommand\xlabel[2][]{\phantomsection\def\@currentlabelname{#1}\label{#2}}
\makeatother

\newcommand{\y}{{\color{Tan}  y}}

\makeatletter
\def\l@subsubsection#1#2{}
\makeatother

\begin{document}
\renewcommand{\theequation}{\arabic{equation}}

\title{The Clifford Hierarchy from Clifford Stabilizers in 2D}

\author{Alison Warman and Sakura Sch\"afer-Nameki}

\affiliation{Mathematical Institute, University of Oxford, Woodstock Road, Oxford, OX2 6GG, United Kingdom}

\begin{abstract}
\noindent The Bravyi-K\"onig theorem limits the constant-depth logical gates of 2D topological Pauli stabilizer codes to be in the Clifford group. Known qubit codes with Clifford stabilizers exceed it by only one level of the Clifford hierarchy, and this was conjectured to hold generally. We show that this bound can be completely bypassed, and that topological codes with Clifford stabilizers in 2D can realize transversal logical gates at any level of the Clifford hierarchy. We present two explicit constructions of codes based on copies of the quantum doubles of $D_4$ and $\Z_2$, with $k$ logical qubits and a logical $C^{k-1}Z$ gate, which is at level $k$ of the Clifford hierarchy. The physical circuits realizing the gates consist only of physical $Z$ or $CZ$ gates. The main new ingredient comes from the logical encoding, which depends crucially on non-Abelian group multiplication. 
\end{abstract}

\maketitle


\paragraf{Introduction.} 
Transversal logical gates are the simplest fault-tolerant gates, since they do not spread errors within a code block. However, the Bravyi-K\"onig (BK) theorem \cite{Bravyi:2012rnv} limits them, and more generally all constant-depth logical gates, in topological 2D Pauli stabilizer codes, to be in the Clifford group. 
In particular, non-Clifford gates that are needed for universal quantum computation require further resources, such as magic states. 
It was recently shown that non-Abelian codes, whose stabilizers are not simply Pauli operators, can evade this restriction and host non-Clifford constant-depth gates. For example, the twisted $\Z_2^3$ gauge theory $D^{\omega_\III}(\Z_2^3)$, or equivalently $D(D_4)$,  \cite{Hu:2012wx,Yoshida:2015cia,Davydova:2025ylx}, has Clifford stabilizers and a constant-depth logical $T=\diag(1,e^{i\pi/4})$ gate in 2D \cite{Kobayashi:2025cfh,Davydova:2025ylx, Warman:2025hov}, and can be implemented \cite{Iqbal:2023wvm} and decoded with a high threshold \cite{Davydova:2025ylx,delaFuente:2026slp}. 
On the other hand, the quantum doubles $D(D_{4N})$ host constant-depth logical $T^{1/N}$ gates \cite{Warman:2025hov} at the cost of more complex stabilizers that include non-Clifford gates \footnote{Alternative recent approaches for non-Clifford gates via non-Abelian codes include lattice surgery with group surface codes \cite{Huang:2025ump,Manjunath:2026swm, Li:2026wly}, 
{non-Abelian quantum LDPC codes \cite{Zhu:2026vec,Christos:2026yyd}, measurements of logical Clifford operators by gauging and code surgery \cite{Hetenyi:2026rbg,Williamson:2026owp,Chang:2026xkz}, and computation with gapped boundaries of quantum doubles \cite{Cong:2016ayp}.}}. 
However, thus far in all known topological qubit codes in 2D, the constant-depth logical gates lie at most one level above the stabilizers, and this was recently conjectured to be a general bound in \cite{Fechisin:2026ywx}.

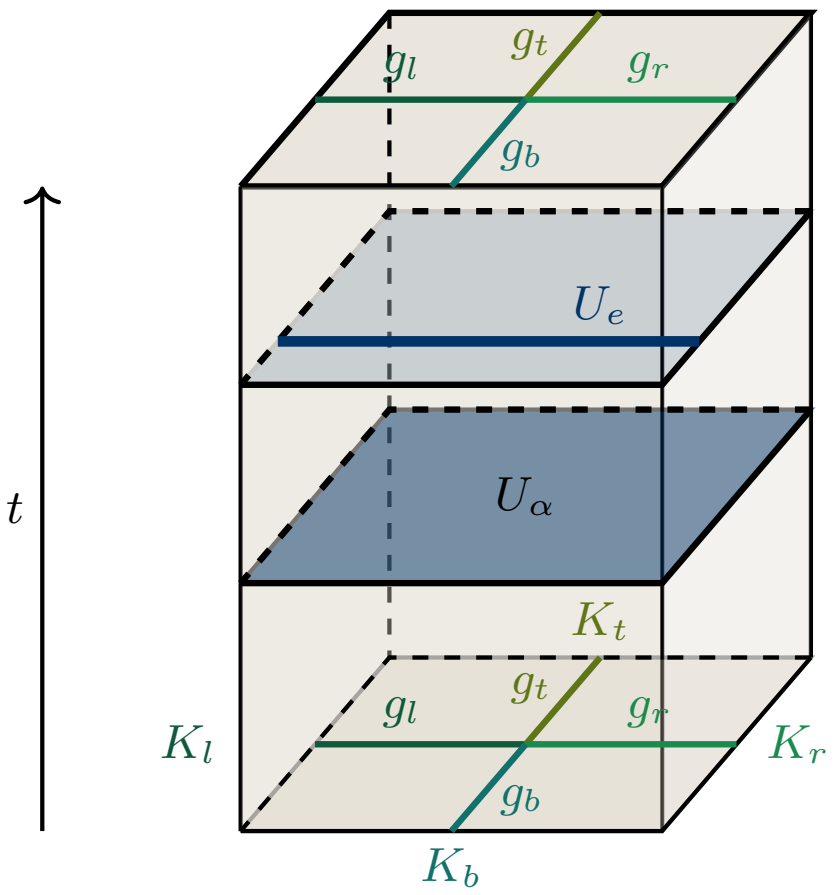
\begin{figure}[t]
\centering
\begin{tikzpicture}
\begin{scope}[shift={(0,0)}, scale=0.82]
\draw [thick, fill= \colFace, opacity=0.5] (0,0) -- (3.4,0) -- (4.6,1.4) -- (1.2,1.4) -- (0,0);
\draw [very thick] (0,0) -- (3.4,0) -- (4.6,1.4) -- (1.2,1.4) -- (0,0);
\draw[thick] (0,0) -- (0,-5.2);
\draw[thick] (3.4,0) -- (3.4,-5.2);
\draw[thick] (4.6,1.4) -- (4.6,1.4-5.2);
\draw[thick, dashed] (1.2,1.4) -- (1.2,1.4-5.2);
\draw[fill= \colFace, opacity=0.4] (0,0) -- (0,-5.2) -- (3.4,-5.2) -- (3.4,0) -- (0,0);
\draw[fill= \colFace, opacity=0.25] (3.4,0) -- (3.4,-5.2) -- (4.6,1.4-5.2) -- (4.6,1.4) -- (3.4,0);
\draw[very thick, \colL] (2.3,0.7) -- (0.6,0.7);
\draw[very thick, \colT] (2.3,0.7) -- (2.9,1.4);
\draw[very thick, \colR] (2.3,0.7) -- (4.0,0.7);
\draw[very thick, \colB] (2.3,0.7) -- (1.7,0);
\node[\colL, above] at (1.3,0.7) {$g_l$};
\node[\colT, left] at (2.65,1.15) {$g_t$};
\node[\colR, above] at (3.3,0.7) {$g_r$};
\node[\colB, right] at (1.95,0.25) {$g_b$};
\begin{scope}[shift={(0,-5.2)}]
\draw [thick, fill= \colFace, opacity=0.3] (0,0) -- (3.4,0) -- (4.6,1.4) -- (1.2,1.4) -- (0,0);
\draw [thick] (0,0) -- (3.4,0) -- (4.6,1.4);
\draw [thick, dashed] (4.6,1.4) -- (1.2,1.4) -- (0,0);
\draw[very thick, \colL] (2.3,0.7) -- (0.6,0.7);
\draw[very thick, \colT] (2.3,0.7) -- (2.9,1.4);
\draw[very thick, \colR] (2.3,0.7) -- (4.0,0.7);
\draw[very thick, \colB] (2.3,0.7) -- (1.7,0);
\node[\colL, above] at (1.3,0.7) {$g_l$};
\node[\colT, left] at (2.65,1.15) {$g_t$};
\node[\colR, above] at (3.3,0.7) {$g_r$};
\node[\colB, right] at (1.95,0.25) {$g_b$};
\node[\colL, left] at (-0.05,0.7) {$K_l$};
\node[\colT, above] at (2.9,1.4) {$K_t$};
\node[\colR, right] at (4.1,0.7) {$K_r$};
\node[\colB, below] at (1.7,0) {$K_b$};
\end{scope}
\begin{scope}[shift={(0,-1.6)}]
\draw [thick, fill= \colU, opacity=0.15] (0,0) -- (3.4,0) -- (4.6,1.4) -- (1.2,1.4) -- (0,0);
\draw [very thick] (0,0) -- (3.4,0) -- (4.6,1.4);
\draw [very thick, dashed] (4.6,1.4) -- (1.2,1.4) -- (0,0);
\draw[\colU, line width=2pt] (0.3,0.35) -- (3.7,0.35);
\node[\colU, above] at (2.9,0.35) {$U_e$};
\end{scope}
\begin{scope}[shift={(0,-3.2)}]
\draw [thick, fill= \colUa, opacity=0.5] (0,0) -- (3.4,0) -- (4.6,1.4) -- (1.2,1.4) -- (0,0);
\draw [very thick] (0,0) -- (3.4,0) -- (4.6,1.4);
\draw [very thick, dashed] (4.6,1.4) -- (1.2,1.4) -- (0,0);
\node at (2.3,0.7) {$U_{\alpha}$};
\end{scope}
\draw[->, thick] (-1.6,-5.2) -- (-1.6,0) node[midway,left] {$t$};
\end{scope}
\end{tikzpicture}
\vspace*{-3mm}
\caption{The quantum double $D(G)$ on a square spatial patch, with time going up. The four vertical faces are the gapped boundaries labeled by the subgroups $K_l,K_t,K_r,K_b\subset G$. A logical state is a configuration of four magnetic anyons with fluxes $g_l,g_t,g_r,g_b$, each ending on a boundary and satisfying $g_lg_t=g_bg_r$. 
The logical gates act on a spatial slice. In Construction I (\colUname), $U_e$ is the line of an Abelian charge anyon between the left and right boundaries, and physically a string of Pauli $Z$ gates. In Construction II (\colUaname), $U_\alpha$ multiplies the state by $\alpha(g_b,g_r)/\alpha(g_l,g_t)$ where $\alpha$ is 
a group 2-cocycle. Both give the transversal logical $C^{k-1}Z$, implemented by physical $Z$ and $CZ$ gates and Clifford stabilizers.
\label{fig:prism}}

\end{figure}

\paragraf{Main Result: Clifford Hierarchy Gates from Clifford Stabilizer Codes.} In this Letter we show that Clifford stabilizers in 2D suffice for transversal logical gates in every level of the Clifford hierarchy. We argue this by providing two explicit constructions, which use the quantum doubles of the non-Abelian groups
\be\ba \label{Gk}
    G_A&=D_4^{\,k_A-1}\times\Z_2 \,, \cr
    G_A\times G_B&=\big(D_4^{\,k_A-1}\times\Z_2\big)\times\big(D_4^{\,k_B-1}\times\Z_2\big) \,.
\ea\ee
Here, $D_4$ is the group of symmetries of a square. The parameters $k_A$ and $k_B$ count the logical qubits of the two factors with $k=k_A+k_B$, when considering the code on a square patch with four gapped boundaries (Fig.~\ref{fig:prism}).
Both constructions act on $k$ logical qubits realizing the logical 
\be \label{CkZ}
    C^{k-1}Z=(-1)^{x_1x_2\cdots x_k}\,,
\ee
which lies in level $k$ of the Clifford hierarchy \cite{Cui:2016bxt}. Multi-controlled logical gates are central to quantum algorithms and to universal gate sets \cite{Shi:2002cli,Aharonov:2003hjn,Gidney:2018haa}, but in a Pauli stabilizer code they require costly magic states. 
With our constructions, once the quantum double codes for \eqref{Gk} are prepared, \eqref{CkZ} is a single layer of one-qubit or two-qubit gates, with stabilizers and physical gates that are all at most Clifford.

How is this possible? In a Pauli stabilizer code a physical Pauli operator that preserves the code space acts on it as a logical Pauli \cite{Gottesman:1997zz}.
However, for a code with Clifford stabilizers this need not hold. 
The logical states are determined from magnetic anyons with fluxes $g\in G$, stretching between the left and right, or top and bottom, boundaries. However, when they cross, non-Abelian group multiplication can force the exponent of a certain group element to be fixed in terms of the others. 

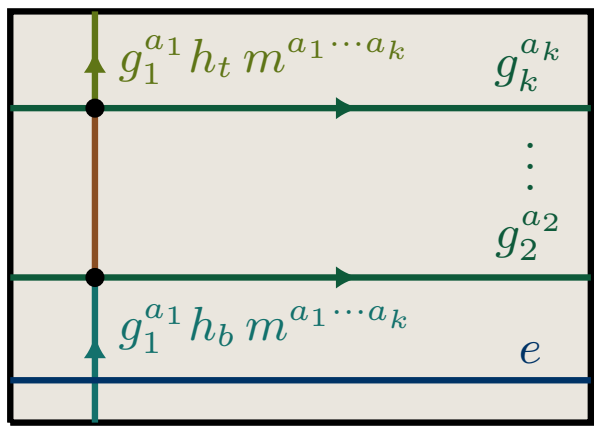
\begin{figure}
$$
\begin{tikzpicture}
\begin{scope}[scale=0.8]
\draw [thick, fill= \colFace, opacity=0.5] (0,0) -- (4.8,0) -- (4.8,3.4) -- (0,3.4) -- (0,0);
\draw [very thick] (0,0) -- (4.8,0) -- (4.8,3.4) -- (0,3.4) -- (0,0);
\draw[very thick, \colL, ->-] (0,1.2) -- (4.8,1.2);
\draw[very thick, \colL, ->-] (0,2.6) -- (4.8,2.6);
\node[\colL, above] at (4.3,1.2) {$g_2^{a_2}$};
\node[\colL] at (4.3,2.25) {$\vdots$};
\node[\colL, above] at (4.3,2.6) {$g_k^{a_k}$};
\draw[very thick, \colB, ->-] (0.7,0) -- (0.7,1.2);
\draw[very thick, \colAux] (0.7,1.2) -- (0.7,2.6);
\draw[very thick, \colT, ->-] (0.7,2.6) -- (0.7,3.4);
\node[\colB, right] at (0.75,0.78) {$g_1^{a_1}h_b\,m^{a_1\cdots a_k}$};
\node[\colT, right] at (0.75,3.0) {$g_1^{a_1}h_t\,m^{a_1\cdots a_k}$};
\draw[very thick, \colU] (0,0.35) -- (4.8,0.35);
\node[\colU, above] at (4.3,0.35) {$e$};
\draw[fill=black] (0.7,1.2) ellipse (0.07 and 0.07);
\draw[fill=black] (0.7,2.6) ellipse (0.07 and 0.07);
\end{scope}
\end{tikzpicture}
$$
\caption{{Action of the charge anyon $e$ (blue) on the logical state: the anyon $e$, which on the lattice is a product of $Z$,  acts on the configuration of anyons that realize the logical state, by a sign $(-1)^{a_1 \cdots a_k}$, which is the $C^{k-1}Z$ gate. In the anyon picture, this arises from $e$ crossing the  $m^{a_1 \cdots a_k}$ anyon. \label{fig:anyonsgen}}}
\end{figure}

Consider a finite group $G$ with a $\Z_2$ factor, and denote by $m$ its magnetic flux.
For a suitable choice of $G$ and boundaries, we show that  a vertical flux $g_1^{a_1}$ that crosses the horizontal fluxes $g_2^{a_2},\dots,g_k^{a_k}$ is forced to carry also the $\Z_2$ flux $m^{a_1\cdots a_k}$, as shown in Fig.~\ref{fig:anyonsgen}: 
the total flux is $g_1^{a_1}h_b\,m^{a_1\cdots a_k}$ below the crossings and $g_1^{a_1}h_t\,m^{a_1\cdots a_k}$ above them, where $h_b$ and $h_t$ are group elements fixed by the boundaries and junctions. The exponent of $m$ is $a_1\cdots a_k$ along the whole vertical path. Then, its corresponding electric (charge) anyon $e\in D(\Z_2)$, which on the lattice is a tensor product of Pauli $Z$ gates and acts on a state with $m$ by a $-1$ phase, will produce $(-1)^{a_1\cdots a_k}=C^{k-1}Z$ when acting on the state in Fig.~\ref{fig:anyonsgen}, resulting in a logical controlled-gate. 

In our first construction, the logical gate $\ol{U}_e=C^{k-1}Z$ is realized by a tensor product of Pauli $Z$ operators along a path between two opposite boundaries, while in the second, $\ol{U}_\alpha=C^{k-1}Z$ is obtained by stacking a group 2-cocycle $\alpha$ as in \cite{Warman:2025hov}, which is implemented as a layer of physical $CZ$ gates on the whole patch. Our setup and the logical gates are illustrated in Fig.~\ref{fig:prism}.

We remark that the non-Clifford nature of the gates does not come from the physical operators implementing them but from the non-Abelian nature of the code, as we explained above. The code stabilizers are Clifford and we expect the constant-depth adaptive protocol of $D(D_4)$ \cite{Tantivasadakarn:2022ceu,Iqbal:2023wvm} to be a natural starting point for its preparation. Then, once our code has been implemented, the logical $C^{k-1}Z$ gate is a depth-one circuit of Pauli $Z$ or $CZ$ gates, which can be applied any number of times during a computation, so we expect our protocols to be favorable for computations that require many non-Clifford gates. We will prove in the End Matter that the Clifford level of our gates cannot decrease by a change of logical basis.

\paragraf{Statement of the Main Theorem.} Let the stabilizers of a $D$-dimensional code lie in $\cC^{(\ell)}$, where $\cC^{(1)}$ is the Pauli group and $V\in\cC^{(\ell+1)}$ if $VPV^{-1}\in\cC^{(\ell)}$ for all Pauli operators $P$ \cite{Gottesman:1999tea}. 
For Pauli stabilizer codes, we have $\ell=1$ and the BK theorem \cite{Bravyi:2012rnv} places the constant-depth logical gates in $\cC^{(D)}$. 
All known qubit codes with stabilizers beyond the Pauli group, listed in Table~\ref{tab:literature}, have constant-depth logical gates in
\be \label{conj}
    \cC^{(\ell+D-1)} \,,
\ee
which was conjectured to be a general bound in \cite{Fechisin:2026ywx}.
We show in this Letter that this conjectured generalized Bravyi-K\"onig bound can be completely evaded: transversal logical gates at arbitrary levels of the hierarchy can be obtained, even in the simplest non-Pauli topological setting with $D=2$ and $\ell=2$.

\begin{table}
$$
\begin{array}{|l|c|c|c|c|}
\hline
\text{Code} & D & \ell & \text{Gate} & \makecell[c]{\text{Level of}\\ \text{gate}} \\
\hline
\text{Pauli \cite{Bravyi:2012rnv}} & D & 1 & & \le D \\
D(D_4)\ \text{\cite{Kobayashi:2025cfh,Davydova:2025ylx, Warman:2025hov}} & 2 & 2 & T & 3 \\
D(D_{4N})\ \text{\cite{Warman:2025hov}} & 2 & 2+\log_2N & T^{1/N} & 3+\log_2N \\
\text{I: } D(G_A) & 2 & 2 & C^{k-1}Z & k \\
\text{II: } D(G_A\times G_B) & 2 & 2 & C^{k-1}Z & k \cr
\hline
\end{array}
$$
\caption{Clifford Level $\ell$ of the stabilizers and of the constant-depth logical gate of topological qubit codes in $D$ dimensions, with $N$ a power of two.
\label{tab:literature}}
\end{table}

\noindent
Our main result is the following: \vspace*{-1mm}
\begin{theorem} \label{thm}
Let $k\ge3$. The following two codes, realized on an $L\times L$ square lattice patch, are qubit codes with Clifford stabilizers, $k$ logical qubits and distance $L$, with a transversal logical $C^{k-1}Z$ gate:

\noindent {\bf Construction I.} The quantum double of $G_A$ in \eqref{Gk} with $k_A=k$ and the boundary subgroups \eqref{KI}, with $3k-2$ physical qubits on each lattice edge. The gate is realized by a tensor product of $L$ single-qubit Pauli $Z$ gates.

\noindent {\bf Construction II.} The quantum double of $G_A\times G_B$ in \eqref{Gk} with $k_A+k_B=k$, $k_A,k_B\ge1$, and the boundary subgroups \eqref{Kgen}, with $3k-4$ physical qubits per lattice edge. The gate is realized as a single layer of $CZ$ gates, transversal between the two code blocks, $D(G_A)$ and $D(G_B)$.
\end{theorem}


\paragraf{Setup.} We place the Kitaev quantum double $D(G)$ \cite{Kitaev:1997wr} on a square, and impose on all its four sides gapped boundary conditions, labeled by subgroups $K_l,K_t,K_r,K_b\le G$ (and trivial 2-cocycles) \cite{Beigi:2010htr}. 
The logical states are ground states of the quantum double Hamiltonian \cite{Kitaev:1997wr,Beigi:2010htr}.
We will characterize them by four magnetic anyons, each ending on a boundary on which its flux condenses, which meet at a common plaquette. 
We denote them by $\ol{|g_l,g_t,g_r,g_b\rangle}$ with $g_l\in K_l$, $g_t\in K_t$, $g_r\in K_r$, $g_b\in K_b$ which satisfy 
\be \label{plaqflat}
g_lg_t=g_bg_r
\ee 
to ensure that the common plaquette has trivial flux (Fig.~\ref{fig:prism}). By choosing a suitable group $G$ as in \eqref{Gk} and boundary conditions, we can encode any number of logical qubits $k$, as we will explain below.

\paragraf{$\bm{D(D_4\times\Z_2)}$ Code.} We begin by discussing the main building block which is the quantum double of the group
\be \label{H}
    G_A=\langle r,s\rangle\times\langle m\rangle\cong D_4\times\Z_2 \,.
\ee
Here, we present the $D_4$ as the symmetries of a square with  rotation $r$ and reflection $s$
\be
D_4=\langle r,s\,|\,r^4=s^2=\id\,,\ srs=r^{-1}\rangle\,,
\ee
while $m$ generates the $\Z_2$ factor and we define
\be \label{rs}
    t:=r^2\,m \,.
\ee
All four elements $sr,s,m,t$ have order two, they satisfy $(sr)\,s=s\,(sr)\,r^2,\;mt=r^2$, and $r^2$, $m$, $t$ are central. 
We place $D(G_A)$ on a square with the boundary subgroups
\be \label{K2}
    K_l=K_r=\langle sr\rangle \,,\quad K_b=\langle s,m\rangle \,,\quad K_t=\langle s,t\rangle \,,
\ee
so that the bottom and top boundaries differ only by $m$ and $t$. 

This setup encodes two logical qubits, whose basis states are $\ol{|a_1,a_2\rangle}$ with $a_1,a_2\in\{0,1\}$.
The horizontal anyon between the left and right boundaries carries the flux $(sr)^{a_2}$, and the vertical anyon between the bottom and top boundaries carries $s^{a_1}$ with further factors that are fixed by the boundary conditions when the two anyons cross. 
For general $a_1,a_2$, the logical state is the following anyon configuration:
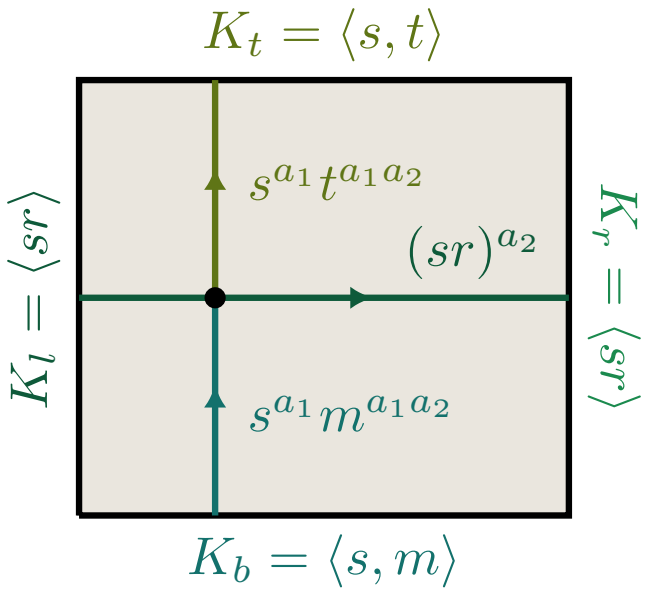
\be \label{anyonsH}
\begin{tikzpicture}[baseline={([yshift=-0.5ex]current bounding box.center)}]
\begin{scope}[scale=0.9]
\draw [thick, fill= \colFace, opacity=0.5] (0,0) -- (3.6,0) -- (3.6,3.2) -- (0,3.2) -- (0,0);
\draw [very thick] (0,0) -- (3.6,0) -- (3.6,3.2) -- (0,3.2) -- (0,0);
\draw[very thick, \colL, ->-] (0,1.6) -- (3.6,1.6);
\draw[very thick, \colB, ->-] (1.0,0) -- (1.0,1.6);
\draw[very thick, \colT, ->-] (1.0,1.6) -- (1.0,3.2);
\node[\colL, above] at (2.9,1.6) {$(sr)^{a_2}$};
\node[\colB, right] at (1.1,0.75) {$s^{a_1}m^{a_1a_2}$};
\node[\colT, right] at (1.1,2.45) {$s^{a_1}t^{a_1a_2}$};
\draw[fill=black] (1.0,1.6) ellipse (0.07 and 0.07);
\node[\colL, rotate=90, above] at (0,1.6) {$K_l=\langle sr\rangle$};
\node[\colR, rotate=-90, above] at (3.6,1.6) {$K_r=\langle sr\rangle$};
\node[\colB, below] at (1.8,0) {$K_b=\langle s,m\rangle$};
\node[\colT, above] at (1.8,3.2) {$K_t=\langle s,t\rangle$};
\end{scope}
\end{tikzpicture}
\ =\ \ol{|a_1,a_2\rangle}
\vspace*{-2mm}
\ee
This forces the exponent of $m$ in $g_b$ to be the product $a_1a_2$ of the two logical bits. To see this, starting with the general bits $a_1$, $a_2$, $A_1$, for $s,sr,m$, respectively, we write
\be \label{g2}
    g_l=g_r=(sr)^{a_2} \,,\quad g_b=s^{a_1}m^{A_1} \,,\quad g_t=s^{a_1}t^{A_1} \,.
\ee
It then follows by straightforward computation that 
\be\ba \label{flat22}
    g_lg_t&=(sr)^{a_2}\,s^{a_1}\,t^{A_1}
    =s^{a_1}\,m^{A_1}\,(sr)^{a_2}\,r^{2(a_1a_2+A_1)} \cr
    &=g_bg_r\,r^{2(a_1a_2+A_1)} \,,
\ea\ee
so the plaquette stabilizer condition \eqref{plaqflat} holds if and only if $A_1=a_1a_2$. The four logical states are therefore:
\be \label{fourstates}
\begin{array}{|c|c|c|c|c|c|}
\hline
a_1 & a_2 & g_l=g_r & g_b & g_t & \ol{U}_e \cr \hline
0 & 0 & \id & \id & \id & +1 \cr
1 & 0 & \id & s & s & +1 \cr
0 & 1 & sr & \id & \id & +1 \cr
1 & 1 & sr & sm & st & -1 \cr \hline
\end{array}
\ee
The last column is the value of the operator $U_e=\prod_{i\in\gamma}Z_{m,i}$ on a  horizontal path $\gamma$: at the logical level, $\ol{U}_e=(-1)^{a_1a_2}=CZ$ for $k=2$.


We now explain how quantum doubles of products of these groups are used for our two constructions of $C^3Z$ and subsequently generalize to any $C^{k-1}Z$.
\paragraf{Construction I: $C^3Z$ from $Z$.} To obtain the first gate beyond the putative bound \eqref{conj} (which is the logical $C^3Z$ gate) from physical Pauli $Z$ gates, we take 
\be 
G_A=D_4^3\times\Z_2
\ee 
with the generators $r_j,s_j$ of the three copies of $D_4$, the generator $m$ of $\Z_2$, and 
\be
t_1=r_1^2s_2\,,\quad t_2=r_2^2s_3\,,\quad t_3=r_3^2m\,.
\ee
The subgroups labeling the four boundaries are
\be\ba \label{KC3Z}
    K_l&=K_r=\langle s_1r_1,s_2r_2,s_3r_3\rangle \,, \cr
    K_b&=\langle s_1,s_2,s_3,m\rangle \,,\quad K_t=\langle s_1,t_1,t_2,t_3\rangle \,,
\ea\ee
and the code has four logical qubits $a_1,\dots,a_4$. Note that the subgroups associated to boundaries that share a corner have trivial intersection, and thus there are no short strings.
A non-trivial $s_1$ flux now crosses $s_1r_1,s_2r_2,s_3r_3$, and each crossing constrains one more generator as in \eqref{anyonsH}, so the boundary holonomies are 
\be\ba \label{gC3Z}
    g_l=g_r&=(s_1r_1)^{a_2}\,(s_2r_2)^{a_3}\,(s_3r_3)^{a_4} \,, \cr
    g_b&=s_1^{a_1}\,s_2^{a_1a_2}\,s_3^{a_1a_2a_3}\,m^{a_1a_2a_3a_4} \,, \cr
    g_t&=s_1^{a_1}\,t_1^{a_1a_2}\,t_2^{a_1a_2a_3}\,t_3^{a_1a_2a_3a_4} \,.
\ea\ee
The string of Pauli $Z$ gates on the qubits for the $\Z_2$ factor $m$ is a $-1$ phase with the exponent of $m$ in $g_b$,
\be\ba \label{WC3Z}
    U_e&=\textstyle\prod_{i\in\gamma}Z_{m,i} \,, \cr
    \ol{U}_e&=(-1)^{a_1a_2a_3a_4}=C^3Z \,,
\ea\ee
which lies in $\cC^{(4)}$ and not in $\cC^{(3)}$. So $L$ single-qubit Pauli gates, on a code with Clifford stabilizers and $10$ qubits per edge (Table~\ref{tab:examples}), already exceed the putative bound \eqref{conj}.

\paragraf{Construction II: $C^3Z$ from $CZ$.} 
We take two copies $G_A\times G_B$ with $G_B=G_A=D_4\times\Z_2$. We write the elements as pairs $(g_A,g_B)$ and the subgroups as products. 
Since all four boundary subgroups are products, $D(G_A\times G_B)$ consists of two decoupled code blocks on the same lattice, $D(G_A)$ with the boundary subgroups \eqref{K2} and $D(G_B)$ with the same boundaries rotated by 90 degrees. 
The boundary subgroups for $D(G_A\times G_B)$ are 
\be \label{K22}
\begin{array}{|c|c|}
\hline
\text{Subgroup} &  \;\;\subset G_A\times G_B \phantom{\subset\;\;} \cr \hline
K_l & \langle sr\rangle\times\langle s,m\rangle \cr
K_r & \langle sr\rangle\times\langle s,t\rangle \cr
K_b & \langle s,m\rangle\times\langle sr\rangle \cr
K_t & \langle s,t\rangle\times\langle sr\rangle \cr \hline
\end{array}
\ee
and the four logical qubits are labeled by $a_1,a_2$ for $G_A$ and $b_1,b_2$ for $G_B$. 

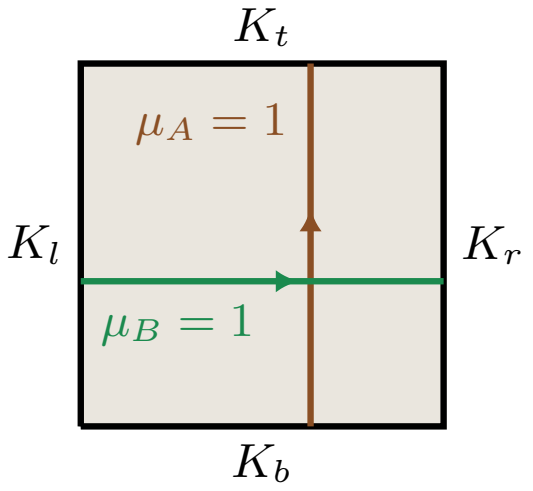
\begin{figure}
\centering
\begin{tikzpicture}
\begin{scope}[scale=0.8]
\draw [thick, fill= \colFace, opacity=0.5] (0,0) -- (3,0) -- (3,3) -- (0,3) -- (0,0);
\draw [very thick] (0,0) -- (3,0) -- (3,3) -- (0,3) -- (0,0);
\draw[very thick, \colAux, ->-] (1.9,0) -- (1.9,3);
\draw[very thick, \colR, ->-] (0,1.2) -- (3,1.2);
\node[\colAux, left] at (1.85,2.5) {$\mu_A=1$};
\node[\colR, below] at (0.8,1.15) {$\mu_B=1$};
\node[left] at (0,1.5) {$K_l$};
\node[right] at (3,1.5) {$K_r$};
\node[below] at (1.5,0) {$K_b$};
\node[above] at (1.5,3) {$K_t$};
\end{scope}
\end{tikzpicture}
\caption{Construction II for $k=4$: the flux lines with $\mu_A=1$ (if $a_1a_2=1$) and $\mu_B=1$ (if $b_1b_2=1$) cross once. Stacking the SPT phase of $\alpha$ in \eqref{alpha} on the spatial slice counts this crossing, so $\ol{U}_\alpha=(-1)^{a_1a_2b_1b_2}=C^3Z$.
\label{fig:anyonsII}}
\end{figure}

Let $\mu_A(g)$ and $\mu_B(g)$ be the exponents of $m$ in $g_A$ and in $g_B$, which are homomorphisms $G_A\times G_B\to\Z_2$. In particular, $\mu_A(t,\id)=\mu_B(\id,t)=1$ since $t=r^2m$, while $\mu_A$ and $\mu_B$ vanish on $sr$.
Their cup product is a group 2-cocycle
\be \label{alpha}
    \alpha(g,h)=(-1)^{\mu_A(g)\,\mu_B(h)} \,,
\ee
whose class is nontrivial (End Matter). On the four boundaries 
\be \label{alphaK}
    \alpha(g,h)=1 \quad \text{for all } g,h\in K_l \,,\ K_t \,,\ K_r \text{ or } K_b \,,
\ee
since $\mu_A$ vanishes on $K_l$ and $K_r$, whose $G_A$ factor is $\langle sr\rangle$, and $\mu_B$ vanishes on $K_b$ and $K_t$ so the gate $U_\alpha$, constructed as in \cite{Warman:2025hov}, does not require any boundary terms. 
With $\mu_A(g_b)=a_1a_2,\; \mu_B(g_r)=b_1b_2,\; \mu_A(g_l)=\mu_B(g_t)=0$,
it evaluates to
\be \label{U22}
    \ol{U}_\alpha=\frac{\alpha(g_b,g_r)}{\alpha(g_l,g_t)}=(-1)^{a_1a_2\,b_1b_2}=C^3Z \,.
\ee
This is the $C^3Z$ gate for the logical bits $a_1,a_2,b_1,b_2$ of $G_A$ and of $G_B$. $\ol{U}_\alpha$ is $-1$ if and only if the flux lines of $m_A$ and of $m_B$ are both present (Fig.~\ref{fig:anyonsII}) and only requires 8 physical qubits per edge (Table~\ref{tab:examples}). 

\paragraf{Constructions I and II for every $k$.} The codes for larger $k$ are obtained by extending one factor at a time, starting from the surface code $G_A=\langle m\rangle\cong\Z_2$ with one logical qubit:
\be \label{extension}
    \langle m\rangle\ \to\ \langle r_1,s_1\rangle\times\langle m\rangle\ \to\ \langle r_1,s_1\rangle\times\langle r_2,s_2\rangle\times\langle m\rangle\ \to\ \cdots
\ee
In the $j$-th step the generator $m$ of the $\Z_2$ factor becomes the reflection $s_j$ of a new copy $\langle r_j,s_j\rangle$ of $D_4$, and a new $\Z_2$ with generator $m$ is added, together with $t_j:=r_j^2m$, which becomes $t_j=r_j^2s_{j+1}$ after the next step. 
After $k_A-1$ steps this gives $G_A$ in \eqref{Gk}. For Construction II, the second group $G_B$ is the same with $k_B$ in place of $k_A$. 

\noindent We define the following subgroups
\be\ba \label{UPQ}
\cR_A&=\langle s_1r_1,\dots,s_{k_A-1}r_{k_A-1}\rangle\,,\cr
\cS_A&=\langle s_1,\dots,s_{k_A-1},m\rangle\,, \cr 
\cT_A&=\langle s_1,t_1,\dots,t_{k_A-1}\rangle\,,
\ea
\ee
and likewise for $G_B$ \footnote{For $k_A=1$: $G_A=\langle m\rangle$, $\cR_A=\{\id\}$, $\cS_A=\cT_A=\langle m\rangle$.}.
The boundary subgroups generalize \eqref{K2} for Construction I,
\be \label{KI}
    K_l=K_r=\cR_A \,,\qquad K_b=\cS_A \,,\qquad K_t=\cT_A \,,
\ee
and \eqref{K22} for Construction II,
\be\ba \label{Kgen}
    K_l&=\cR_A\times\cS_B \,,\qquad K_r=\cR_A\times\cT_B \,, \cr
    K_b&=\cS_A\times\cR_B \,,\qquad K_t=\cT_A\times\cR_B \,.
\ea\ee
So each step adds $s_jr_j$ to the left and right boundaries, $m$ to the bottom boundary and $t_j$ to the top boundary. 

The factor $G_A$ encodes $k_A$ logical qubits, specified by the exponent $a_1$ of $s_1$ and the exponents $a_{j+1}$ of $s_jr_j$. 
Similarly to \eqref{gC3Z}, the boundary holonomies are
\be\ba \label{gtower}
    g_l&=g_r=(s_1r_1)^{a_2}\cdots(s_{k_A-1}r_{k_A-1})^{a_{k_A}} \,, \cr
    g_b&=s_1^{A_0}\,s_2^{A_1}\cdots s_{k_A-1}^{A_{k_A-2}}\,m^{A_{k_A-1}} \,, \cr
    g_t&=s_1^{A_0}\,t_1^{A_1}\cdots t_{k_A-1}^{A_{k_A-1}} \,,
\ea\ee
with $A_0=a_1$ and $A_j=a_1a_2\cdots a_{j+1}$. The exponents of $g_l=g_r$ and the exponent of $s_1$ are the logical bits, while all other exponents are fixed by the condition $g_lg_t=g_bg_r$.

In Construction I, as in \eqref{WC3Z}, the $e$ anyon acting on the logical state with $m^{a_1\cdots a_k}$ has the physical and logical gate decompositions:
\be\ba \label{Wgen}
    U_e&=\textstyle\prod_{i\in\gamma}Z_{m,i} \,, \cr
    \ol{U}_e&=(-1)^{a_1a_2\cdots a_k}=C^{k-1}Z \,.
\ea\ee
In Construction II, with $\mu_A$ and $\mu_B$ the exponents of $m$ in $g_A$ and in $g_B$, the cocycle is again \eqref{alpha}, it still satisfies \eqref{alphaK}, and the gate is
\be \label{Ugen}
    \ol{U}_\alpha=\frac{\alpha(g_b,g_r)}{\alpha(g_l,g_t)}=(-1)^{a_1\cdots a_{k_A}b_1\cdots b_{k_B}}=C^{k-1}Z \,.
\ee

\begin{table}
$$
\begin{array}{|c|c|c|c|}
\hline
k & \text{Gate} & \makecell[c]{\text{I: } G_A\\ \text{(qubits per edge)}} & \makecell[c]{\text{II: } G_A\times G_B\\ \text{(qubits per edge)}} \\
\hline
3 & CCZ & D_4^2\times\Z_2\ \quad (7) & (D_4\times\Z_2)\times\Z_2\  \quad (5) \\
4 & C^3Z & D_4^3\times\Z_2\ \quad (10) & (D_4\times\Z_2)\times(D_4\times\Z_2)\ (8) \\
k & C^{k-1}Z & D_4^{k-1}\times\Z_2\  (3k-2) & G_A\times G_B\ \quad (3k-4) \cr
\hline
\end{array}
$$
\caption{
Constructions I and II  of the logical $C^{k-1}Z$ gate on $k$ logical qubits, with the number of physical qubits on each edge of the lattice.
\label{tab:examples}}
\end{table}

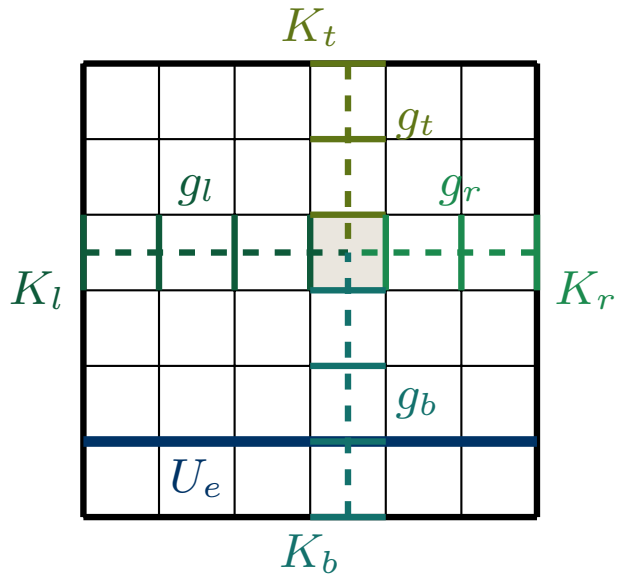
\begin{figure}[t]
\centering
\begin{tikzpicture}
\begin{scope}[shift={(0,0)}, scale=1]
\draw [fill= \colFace, opacity=0.5, draw=none] (1.5,1.5) -- (2,1.5) -- (2,2) -- (1.5,2) -- (1.5,1.5);
\draw[step=0.5,black] (0,0) grid (3,3);
\draw[black, very thick] (0,3) -- (3,3);
\draw[black, very thick] (3,3) -- (3,0);
\draw[black, very thick] (0,3) -- (0,0);
\draw[black, very thick] (0,0) -- (3,0);
\node[\colL, left] at (0,1.5) {$K_l$};
\node[\colT, above] at (1.5,3) {$K_t$};
\node[\colR, right] at (3,1.5) {$K_r$};
\node[\colB, below] at (1.5,0) {$K_b$};
\draw[\colU, line width=2pt] (0,0.5) -- (3,0.5);
\node[\colU, below] at (0.75,0.5) {$U_e$};
\draw[\colL, very thick, dashed] (0,1.75) -- (1.75,1.75);
\draw[\colL, very thick] (0,1.5) -- (0,2);
\draw[\colL, very thick] (0.5,1.5) -- (0.5,2);
\draw[\colL, very thick] (1,1.5) -- (1,2);
\draw[\colL, very thick] (1.5,1.5) -- (1.5,2);
\draw[\colT, very thick, dashed] (1.75,3) -- (1.75,1.75);
\draw[\colT, very thick] (2,3) -- (1.5,3);
\draw[\colT, very thick] (2,2.5) -- (1.5,2.5);
\draw[\colT, very thick] (2,2) -- (1.5,2);
\draw[\colR, very thick, dashed] (3,1.75) -- (1.75,1.75);
\draw[\colR, very thick] (2,1.5) -- (2,2);
\draw[\colR, very thick] (2.5,1.5) -- (2.5,2);
\draw[\colR, very thick] (3,1.5) -- (3,2);
\draw[\colB, very thick] (2,1.5) -- (1.5,1.5);
\draw[\colB, very thick] (2,1) -- (1.5,1);
\draw[\colB, very thick] (2,0.5) -- (1.5,0.5);
\draw[\colB, very thick] (2,0) -- (1.5,0);
\draw[\colB, very thick, dashed] (1.75,0) -- (1.75,1.75);
\node[\colL, above] at (0.75,1.95) {$g_l$};
\node[\colT, right] at (1.95,2.6) {$g_t$};
\node[\colR, above] at (2.5,1.95) {$g_r$};
\node[\colB, right] at (1.95,0.75) {$g_b$};
\end{scope}
\end{tikzpicture}
\caption{The codes on a square lattice ($L=6$).  $K_l,K_t,K_r,K_b$ label the gapped boundaries given in \eqref{KI}, for Construction I. A logical state is created by four magnetic anyons with fluxes $g_l,g_t,g_r,g_b$ and $g_lg_t=g_bg_r$ to ensure trivial flux at their common plaquette $p_*$ (shaded). The gate $U_e$ of Construction I is a string of single-qubit Pauli $Z$ gates on the edges of a line between the left and right boundaries (blue). It crosses the flux $g_b$ and detects its exponent of $m$, and its logical action is the same when crossing $g_t$. 
\label{fig:lattice}}
\end{figure}

Further details and proofs are given in the End Matter, and the groups, the 2-cocycles and the logical gates can also be checked for given $k$ with the GAP functions in \texttt{AlgebrasTQD} \cite{Warman:2026eml}.

\paragraf{Lattice Model and Hamiltonian.} We use Kitaev's quantum double model $D(G)$ \cite{Kitaev:1997wr} on an $L\times L$ square lattice, with boundaries \cite{Beigi:2010htr,Albert:2021vts}. Every edge $i$ carries $\bC[G]$ with basis $|g_i\rangle$, $g_i\in G$, and the Hamiltonian is
\be\ba \label{Ham}
H=-\sum_vA_v-\sum_pB_p-\sum_{i\subset\partial}B_i \,,\quad  
A_v=\frac{1}{|K_v|}\sum_{g\in K_v}A_v^g \,,
\ea\ee
where $K_v$ is $G$ in the bulk, or a boundary subgroup on the boundary, and $K_v=\{\id\}$ at the four corners, where adjacent subgroups intersect trivially. 
The vertex operator $A_v^g$ applies $L^g|h\rangle=|gh\rangle$ to the edges that exit $v$ and $R^g|h\rangle=|hg^{-1}\rangle$ to those that enter $v$, whose circuits we show in Fig.~\ref{fig:circuits}(a). 
The projector $B_p$ imposes \eqref{plaqflat} on the plaquette $p$, and $B_i$ imposes $g_i\in K$ on an edge $i$ of a boundary with subgroup $K$. 

The logical states are created by four magnetic ribbon operators \cite{Kitaev:1997wr,Beigi:2010htr,Bombin:2007qv} \footnote{We can use this simplified expression in terms of tensor products of left-multiplication operators on each ribbon $\xi$ because we are acting on identity group elements.}
\be \label{logstate}
    \ol{|g_l,g_t,g_r,g_b\rangle}=\prod_vA_v\;L^{g_l}_{\xi_l}L^{g_t}_{\xi_t}L^{g_r}_{\xi_r}L^{g_b}_{\xi_b}\bigotimes_i|\id\rangle_i \,,
\ee
that meet at a plaquette $p_*$ (Fig.~\ref{fig:lattice}).

\paragraf{Qubit Realization.}
{Both codes can be realized in terms of physical qubits, where for $D(D_4)$, we map a basis state for each lattice edge onto three qubits as $|r^\rho s^\sigma\rangle\mapsto|\rho_1\rho_0\sigma\rangle$, where $\rho=2\rho_1+\rho_0$ \cite{Warman:2025hov}. This gives 
$3k_A-2$ physical qubits per edge for $G_A$ and $3k-4$ for $G_A\times G_B$.
The details of the stabilizer and gate qubit circuits can be found in the End Matter. On the qubits of the $j$-th copy of $D(D_4)$, the element $r_j^{\rho_j}s_j^{\sigma_j}$ with $\rho_j=2\rho_{j,1}+\rho_{j,0}$ is $(s_jr_j)^{\rho_{j,0}}\,s_j^{\rho_{j,0}+\sigma_j}$ up to a power of $r_j^2$. So the exponents of $s_jr_j$ and of $s_j$ are the bits $\rho_{j,0}$ and $\rho_{j,0}+\sigma_j$, while the exponent of $m$ on each site is stored in its qubit. The lattice realizations of the charge anyons that detect these exponents are, on a path $\gamma$
\be\ba \label{Zstrings}
    Z_{s_j}(\gamma)&=\textstyle\prod_{i\in\gamma}(\bbI\otimes Z\otimes Z)_{j,i} \,, \cr
    Z_{s_jr_j}(\gamma)&=\textstyle\prod_{i\in\gamma}(\bbI\otimes Z\otimes\bbI)_{j,i} \,, \cr
    Z_m(\gamma)&=\textstyle\prod_{i\in\gamma}Z_{m,i} \,.
\ea\ee
Denoting by $\gamma_h$ a horizontal path between the left and right boundaries and by $\gamma_v$ a vertical path between the bottom and top boundaries, the strings $Z_{s_j}(\gamma_h)$, $Z_m(\gamma_h)$ and $Z_{s_jr_j}(\gamma_v)$ commute with the code stabilizers, and from \eqref{gtower} their action on the logical state $\ol{|a_1,\dots,a_{k_A}\rangle}$ is
\be\ba \label{Zbar}
    Z_{s_1}(\gamma_h)&=(-1)^{a_1}={\ol{Z}_1} \,, \cr
    Z_{s_jr_j}(\gamma_v)&=(-1)^{a_{j+1}}={\ol{Z}_{j+1}} \,, \cr
    Z_{s_{j+1}}(\gamma_h)&=(-1)^{a_1a_2\cdots a_{j+1}}=C^jZ \,, \cr
    Z_m(\gamma_h)&=(-1)^{a_1a_2\cdots a_{k_A}}=C^{k_A-1}Z \,,
\ea\ee
with $1\le j\le k_A-1$ in the second line and $1\le j\le k_A-2$ in the third line. The first two operators are the logical $\ol{Z}$ operators of the $k_A$ logical qubits, while the last two contain the whole tower $CZ,CCZ,\dots,C^{k_A-1}Z$, where $C^jZ$ acts on the first $j+1$ logical qubits, each as a string of physical Pauli $Z$ gates, and the last one is $U_e$ in \eqref{Wgen}. 

The logical $\ol{X}$ operators, which flip the logical bits, are instead constructed from the magnetic ribbon operators, similarly to \eqref{logstate}. They are not Pauli $X$ strings, since by \eqref{gtower} flipping a logical bit also changes the exponents $A_1,\dots,A_{k_A-1}$ in $g_b$ and $g_t$ by products of the other logical bits.

}

\paragraf{Logical Gates.} The gate of Construction I is (Fig.~\ref{fig:circuits}(c))
\be \label{Ucircuit}
U_e=\textstyle\prod_{x=0}^{L-1}Z_{m,(x,y)} \,,
\ee
on the qubits of $m$ on a horizontal path. It commutes with all stabilizers, since $A_v^g$ applies $X$ to the qubits of $m$ of the two edges on the row at $v$ or leaves them unchanged, and the exponent of $m$ vanishes on the subgroup $\cR_A$ of the left and right boundaries. 

The gate of Construction II is $U_\alpha=\prod_pM^\alpha_p$, where $M^\alpha_p$ evaluates $\alpha$ on each triangulated plaquette $p$ as \cite{Warman:2025hov} \vspace*{-2mm}
\be \label{Ualpha}
\begin{split}
\begin{tikzpicture}
\begin{scope}[shift={(-0.9,0)}, scale= 0.8]
    \draw[->-] (-1,-0.5) to (-1,0.5);
    \node at (-1.3,0) {${g_l}$};
    \draw[->-] (-1,0.5) to (0,0.5);
    \node at (-0.5,0.8) {${g_t}$};
    \draw[->-] (0,-0.5) to (0,0.5);
    \node at (0.3,0) {${g_r}$};
    \draw[->-] (-1,-0.5) to (0,-0.5);
    \node at (-0.5,-0.8) {${g_b}$};
    \draw  (-1,-0.5) -- (0,0.5) ;
    \node at (-1.9,0) {$M^\alpha_p\, \Bigg|$};
    \node[right] at (0.3, 0) {$\Biggl>$};
    \node[right] at (0.75,0) {$=\ \dfrac{\alpha(g_b,g_r)}{\alpha(g_l,g_t)}\Bigg|$};
\end{scope}
\begin{scope}[shift={(3,0)}, scale= 0.8]
    \draw[->-] (-1,-0.5) to (-1,0.5);
    \node at (-1.3,0) {${g_l}$};
    \draw[->-] (-1,0.5) to (0,0.5);
    \node at (-0.5,0.8) {${g_t}$};
    \draw[->-] (0,-0.5) to (0,0.5);
    \node at (0.3,0) {${g_r}$};
    \draw[->-] (-1,-0.5) to (0,-0.5);
    \node at (-0.5,-0.8) {${g_b}$};
    \draw  (-1,-0.5) -- (0,0.5) ;
    \node[right] at (0.3, 0) {$\Biggl>$};
\end{scope}
\end{tikzpicture}
\end{split}
\vspace*{-2mm}
\ee
The phase it produces is the same on all configurations of a logical state (End Matter), which gives \eqref{Ugen}. 

By \eqref{alpha}, $M^\alpha_p$ only involves the qubits $m_A$ and $m_B$, so on the lattice $U_\alpha=\prod_p M_p^\alpha$ is realized as
\be \begin{split} \label{Ualphacircuit}
U_\alpha=\textstyle\prod_pCZ_{(m_A,b),(m_B,r)}\,CZ_{(m_A,l),(m_B,t)} \,,
\end{split}\ee
where $CZ_{(m_A,b),(m_B,r)}$ is the $CZ$ gate of the qubit $m_A$ of the bottom edge $b$ and the qubit $m_B$ of the right edge $r$ of $p$, and likewise for the left and top edges. As the transversal $CZ$ gate between two surface codes, it maps every vertex stabilizer to itself times plaquette $Z$ checks of the other code block (End Matter), so it preserves the code space. We present its circuit decomposition in Fig.~\ref{fig:circuits}(d). Finally, $\ol{U}_e$ and $\ol{U}_\alpha$ are both $C^{k-1}Z$, which is at level $k$ of the Clifford hierarchy, which completes the proof of our Theorem.


\paragraf{Fault Tolerance and Outlook.} We have shown that Clifford stabilizers in 2D suffice for transversal logical gates in every level of the Clifford hierarchy, where the level is set by the non-Abelian group and the boundary conditions. Both our gates have depth one: $U_e$ is a product of single-qubit gates, while a single fault in $U_\alpha$ affects at most two qubits of one plaquette, one in each code block. In the bulk, our codes are stacks of surface codes $D(\Z_2)$ and of $D(D_4)$, which, despite being non-Abelian, can be decoded with a high threshold by the just-in-time decoder \cite{Davydova:2025ylx,delaFuente:2026slp}, and in this setting also the protocol of \cite{Warman:2026flz} simplifies substantially. Natural next steps are to extend these decoders to the boundaries and to the code switching, and a threshold analysis of the full protocol.

\newpage 
\clearpage
\normalsize
\section*{End Matter}

We now provide some more mathematical and implementation details for our constructions.

\paragraf{The 2-Cocycle.} Here, we prove that \eqref{alpha} is a 2-cocycle in a non-trivial cohomology class and provide further details on Construction II of Theorem~\ref{thm}. We denote by $\mu_A$ and $\mu_B$ the projections of $G_A\times G_B$ onto the direct factors $\langle m\rangle\cong\Z_2$ of $G_A$ and of $G_B$, so modulo $2$
\be\ba \label{muhom}
    \mu_A(gh)&=\mu_A(g)+\mu_A(h) \,, \cr
    \mu_B(gh)&=\mu_B(g)+\mu_B(h) \,.
\ea\ee
By \eqref{alpha} and \eqref{muhom}, the two sides of the cocycle condition
\be \label{cocycle}
    \alpha(g_1,g_2)\,\alpha(g_1g_2,g_3)=\alpha(g_1,g_2g_3)\,\alpha(g_2,g_3)
\ee
are $(-1)$ to the powers
\be\ba \label{cocycleproof}
    &\mu_A(g_1)\mu_B(g_2)+\big(\mu_A(g_1)+\mu_A(g_2)\big)\mu_B(g_3) \,, \cr
    &\mu_A(g_1)\big(\mu_B(g_2)+\mu_B(g_3)\big)+\mu_A(g_2)\mu_B(g_3) \,,
\ea\ee
which are equal. So \eqref{cocycle} holds for all $g_1,g_2,g_3\in G_A\times G_B$, and since $\mu_A(\id)=\mu_B(\id)=0$, $\alpha$ is normalized, $\alpha(\id,g)=\alpha(g,\id)=1$. For commuting $g,h$, a coboundary $\beta(g)\beta(h)/\beta(gh)$ is symmetric, so the ratio $\alpha(g,h)/\alpha(h,g)$ depends only on the class of $\alpha$, and
\be \label{nontrivial}
    \frac{\alpha(g,h)}{\alpha(h,g)}=\frac{-1}{+1}=-1 \quad \text{for } g=(m,\id) \,,\ h=(\id,m) \,.
\ee
So the class of $\alpha$ in $H^2(G_A\times G_B,U(1))$ is nontrivial. By \eqref{UPQ}-\eqref{Kgen}, $\mu_A|_{K_l}=\mu_A|_{K_r}=\mu_B|_{K_b}=\mu_B|_{K_t}=0$, so $\alpha$ is identically 1 on the boundaries, i.e. \eqref{alphaK} for every $k_A,k_B\ge1$.

\paragraf{The Logical Gate.} We evaluate the 2-cocycle stacking circuit of \cite{Warman:2025hov}, for a 2-cocycle $\alpha$ whose restriction to each boundary subgroup $K$ is identically one, so all boundary functions $\beta_K$ can be taken to be 1. We evaluate it on the representative of \eqref{logstate}, as in Fig.~\ref{fig:lattice}. Since $\alpha$ is normalized,
\be\ba \label{Mpvalues}
    \alpha(\id,g_l)/\alpha(g_l,\id)=
    \alpha(g_b,\id)/\alpha(\id,g_b)=
    \alpha(\id,\id)=1 \,,
\ea\ee
so only $p_*$ contributes. 
By \eqref{gtower} for $G_A$, and for $G_B$ with the two pairs of opposite boundaries exchanged, its four edges carry
\be\ba \label{gfull}
    g_l&=\big(\textstyle\prod_j(s_jr_j)^{a_{j+1}}\,,\ \textstyle\prod_js_j^{B_{j-1}}\,m^{B_{k_B-1}}\big) \,, \cr
    g_r&=\big(\textstyle\prod_j(s_jr_j)^{a_{j+1}}\,,\ s_1^{b_1}\textstyle\prod_jt_j^{B_j}\big) \,, \cr
    g_b&=\big(\textstyle\prod_js_j^{A_{j-1}}\,m^{A_{k_A-1}}\,,\ \textstyle\prod_j(s_jr_j)^{b_{j+1}}\big) \,, \cr
    g_t&=\big(s_1^{a_1}\textstyle\prod_jt_j^{A_j}\,,\ \textstyle\prod_j(s_jr_j)^{b_{j+1}}\big) \,,
\ea\ee
with $B_j=b_1b_2\cdots b_{j+1}$. Denoting by $A:=A_{k_A-1}=a_1\cdots a_{k_A}$, $B:=B_{k_B-1}=b_1\cdots b_{k_B}$ and $t_{k_A-1}=r_{k_A-1}^2m$, we have
\be \label{muvalues}
\begin{array}{|c|c|c|c|c|}
\hline
 & g_l & g_t & g_r & g_b \cr \hline
\mu_A & 0 & A & 0 & A \cr
\mu_B & B & 0 & B & 0 \cr \hline
\end{array}
\vspace*{1mm}
\ee
which inserted into \eqref{alpha} gives
\be\ba \label{Ucalc}
    \alpha(g_b,g_r)&=(-1)^{\mu_A(g_b)\,\mu_B(g_r)}=(-1)^{AB} \,, \cr
    \alpha(g_l,g_t)&=(-1)^{\mu_A(g_l)\,\mu_B(g_t)}=1 \,,
\ea\ee
resulting in
\be \label{Uproof}
    \ol{U}_\alpha=\frac{\alpha(g_b,g_r)}{\alpha(g_l,g_t)}=(-1)^{a_1a_2\cdots a_{k_A}\,b_1b_2\cdots b_{k_B}}=C^{k-1}Z \,.
\ee
In \eqref{Ualphacircuit} the qubit $m_A$ of a horizontal (vertical) edge enters only the gate of the plaquette above (to the right of) it, and the qubit $m_B$ only that of the plaquette below (to the left of) it, so $U_\alpha$ is a depth-one layer of $CZ$ gates on disjoint pairs of qubits. If the qubit $m_A$ of the edge $i$ is paired with the qubit $m_B$ of the edge $i'$, then
\be\ba \label{UalphaX}
    U_\alpha X_{m_A,i}U_\alpha^\dagger&=X_{m_A,i}\,Z_{m_B,i'} \,, \cr
    U_\alpha X_{m_B,i'}U_\alpha^\dagger&=X_{m_B,i'}\,Z_{m_A,i} \,,
\ea\ee
while $U_\alpha$ commutes with the gates on all other qubits. Since the product of $Z_{m}$ operators around a plaquette is a stabilizer, $U_\alpha$ preserves the stabilizer group. A similar computation can be performed for the boundaries. Since it preserves the logical space, the phase is independent of the physical state, thus
\be \label{plaqgate}
    \ol{U}_\alpha\;\ol{|g_l,g_t,g_r,g_b\rangle}=\frac{\alpha(g_b,g_r)}{\alpha(g_l,g_t)}\;\ol{|g_l,g_t,g_r,g_b\rangle} \,.
\ee

\paragraf{Code Switching.} Similarly to \cite{Warman:2025hov}, we can perform code-switching by condensing the $r^2$ anyon of each $D(D_4)$ copy. For the codes of this Letter, this results in $D(\Z_2)$ surface codes of which $k$ carry the logical qubits. The boundary subgroups meet $\langle r_j^2\rangle$ trivially, so their images define $D(\Z_2)$ surface codes, one for each of the generators $s_jr_j$, $s_j$ and $m$.
On the lattice, this map is realized by measuring all qubits $\rho_{j,1}$ in the $X$ basis and post-selecting the $+1$ outcomes. 
In the other direction, we initialize the qubits $\rho_{j,1}$ in $|+\rangle$ and project onto the $+1$ eigenstates of the stabilizers of $D(G_A)$ or $D(G_A\times G_B)$. 

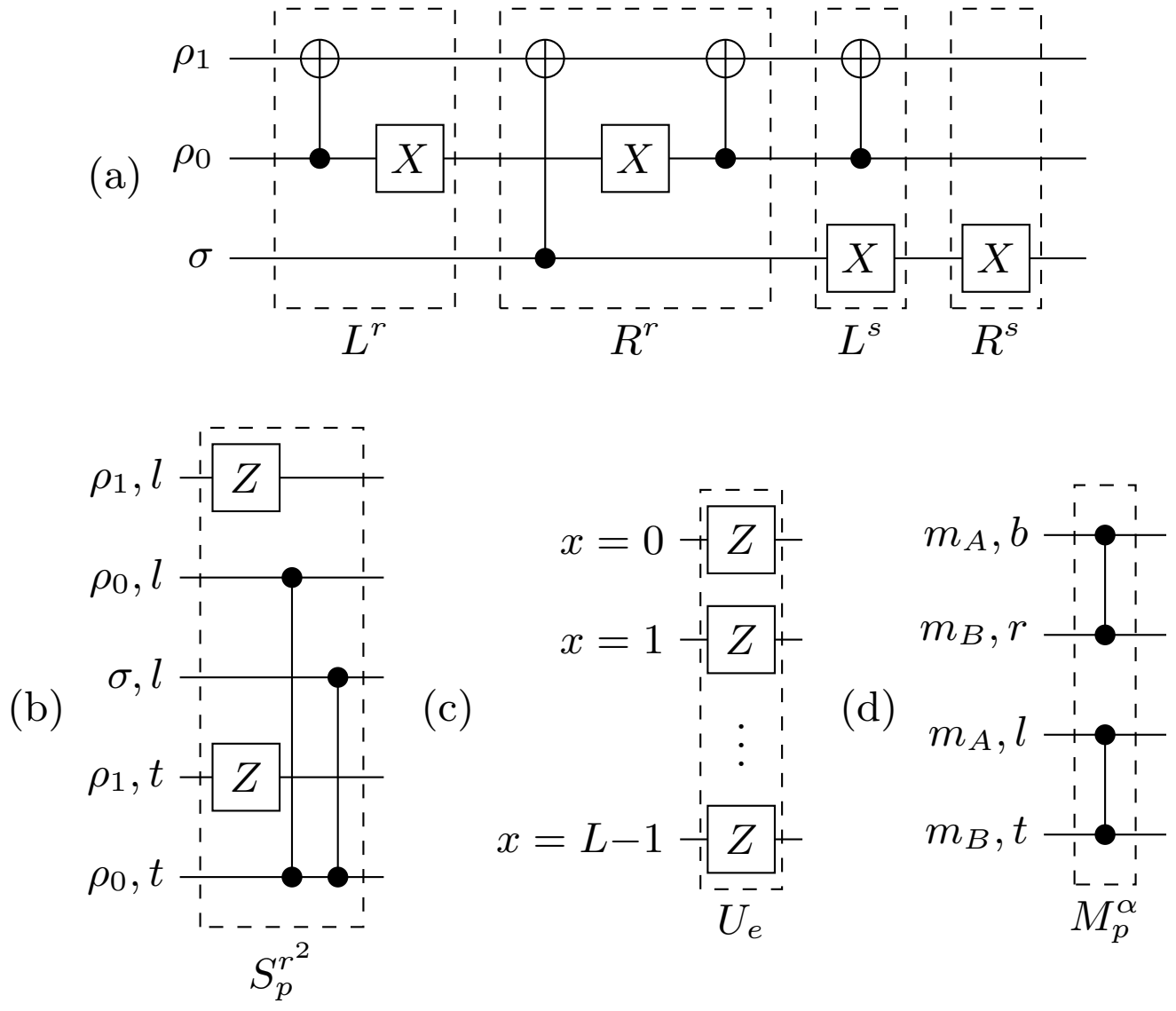
\begin{figure}[t!]
\centering
\tikzset{targ/.style={circle,draw,fill=white,minimum size=8pt,inner sep=0pt,path picture={\draw (path picture bounding box.north)--(path picture bounding box.south) (path picture bounding box.west)--(path picture bounding box.east);}},
gbox/.style={draw,fill=white,minimum size=0.5cm,inner sep=1pt}}
(a)\ {\begin{tikzpicture}[x=0.42cm,y=0.62cm,font=\small,baseline=(current bounding box.center)]
\foreach \y in {0,1.2,2.4}{\draw (0,\y)--(15.2,\y);}
\node[left] at (0,2.4) {$\rho_1$}; \node[left] at (0,1.2) {$\rho_0$}; \node[left] at (0,0) {$\sigma$};
\draw (1.6,1.2)--(1.6,2.4); \fill (1.6,1.2) circle (2.2pt); \node[targ] at (1.6,2.4) {};
\node[gbox] at (3.2,1.2) {$X$};
\draw[dashed] (0.8,-0.6) rectangle (4,3); \node[below] at (2.4,-0.6) {$L^r$};
\draw (5.6,0)--(5.6,2.4); \fill (5.6,0) circle (2.2pt); \node[targ] at (5.6,2.4) {};
\node[gbox] at (7.2,1.2) {$X$};
\draw (8.8,1.2)--(8.8,2.4); \fill (8.8,1.2) circle (2.2pt); \node[targ] at (8.8,2.4) {};
\draw[dashed] (4.8,-0.6) rectangle (9.6,3); \node[below] at (7.2,-0.6) {$R^r$};
\draw (11.2,1.2)--(11.2,2.4); \fill (11.2,1.2) circle (2.2pt); \node[targ] at (11.2,2.4) {};
\node[gbox] at (11.2,0) {$X$};
\draw[dashed] (10.4,-0.6) rectangle (12,3); \node[below] at (11.2,-0.6) {$L^s$};
\node[gbox] at (13.6,0) {$X$};
\draw[dashed] (12.8,-0.6) rectangle (14.4,3); \node[below] at (13.6,-0.6) {$R^s$};
\end{tikzpicture}}

\vspace{0.4cm}
\noindent(b)\,{\begin{tikzpicture}[x=0.38cm,y=0.62cm,font=\small,baseline=(current bounding box.center)]
\foreach \y in {0,1.2,2.4,3.6,4.8}{\draw (-0.4,\y)--(3.6,\y);}
\node[left] at (-0.4,4.8) {$\rho_1,l$}; \node[left] at (-0.4,3.6) {$\rho_0,l$}; \node[left] at (-0.4,2.4) {$\sigma,l$};
\node[left] at (-0.4,1.2) {$\rho_1,t$}; \node[left] at (-0.4,0) {$\rho_0,t$};
\node[gbox] at (0.9,4.8) {$Z$}; \node[gbox] at (0.9,1.2) {$Z$};
\draw (1.8,0)--(1.8,3.6); \fill (1.8,0) circle (2.2pt); \fill (1.8,3.6) circle (2.2pt);
\draw (2.7,0)--(2.7,2.4); \fill (2.7,0) circle (2.2pt); \fill (2.7,2.4) circle (2.2pt);
\draw[dashed] (0,-0.6) rectangle (3.2,5.4); \node[below] at (1.6,-0.6) {$S_p^{r^2}$};
\end{tikzpicture}}%
\hfill%
(c)\,{\begin{tikzpicture}[x=0.38cm,y=0.62cm,font=\small,baseline=(current bounding box.center)]
\foreach \y in {0,2.4,3.6}{\draw (0,\y)--(2.4,\y);}
\node[left] at (0,3.6) {$x=0$}; \node[left] at (0,2.4) {$x=1$};
\node[left] at (0,0) {$x=L{-}1$}; \node at (1.2,1.3) {$\vdots$};
\node[gbox] at (1.2,3.6) {$Z$}; \node[gbox] at (1.2,2.4) {$Z$}; \node[gbox] at (1.2,0) {$Z$};
\draw[dashed] (0.4,-0.6) rectangle (2,4.2); \node[below] at (1.2,-0.6) {$U_e$};
\end{tikzpicture}}%
\hfill%
(d)\,{\begin{tikzpicture}[x=0.38cm,y=0.62cm,font=\small,baseline=(current bounding box.center)]
\foreach \y in {0,1.2,2.4,3.6}{\draw (0,\y)--(2.4,\y);}
\node[left] at (0,3.6) {$m_A,b$}; \node[left] at (0,2.4) {$m_B,r$};
\node[left] at (0,1.2) {$m_A,l$}; \node[left] at (0,0) {$m_B,t$};
\draw (1.2,2.4)--(1.2,3.6); \fill (1.2,2.4) circle (2.2pt); \fill (1.2,3.6) circle (2.2pt);
\draw (1.2,0)--(1.2,1.2); \fill (1.2,0) circle (2.2pt); \fill (1.2,1.2) circle (2.2pt);
\draw[dashed] (0.6,-0.6) rectangle (1.8,4.2); \node[below] at (1.2,-0.6) {$M^\alpha_p$};
\end{tikzpicture}}
\caption{
Circuits of the stabilizers and logical gates: A dot connected to $\oplus$ is a CNOT gate, with the dot on the control qubit, and two connected dots are a $CZ$ gate. {\bf (a)} The gates $L^g$ and $R^g$ of \cite{Warman:2025hov} for the generators $r$ and $s$ of a copy of $D_4$ on its three qubits $\rho_1,\rho_0,\sigma$. 
{\bf (b)} The plaquette stabilizer $S_p^{r^2}$ in \eqref{Sp} of a copy of $D_4$ on the qubits of the left and top edges $l,t$ of $p$, which acts in the same way on the bottom and right edges $b,r$. The other plaquette stabilizers $(S_p^r)^2$, $(S_p^s)^2$ and $Z_m(\partial p)$ are products of single-qubit Pauli $Z$ gates on the four edges of $p$. {\bf (c)} The gate $U_e$ of Construction I, a Pauli $Z$ gate on the qubit $m$ of each horizontal edge of a row. {\bf (d)} The plaquette gate for $U_\alpha=\prod_pM^\alpha_p$ of Construction II in \eqref{Ualphacircuit}: two $CZ$ gates between the qubits $m_A$ and $m_B$ of the bottom and right edges and of the left and top edges of $p$. Both gate circuits have depth one.
\label{fig:circuits}}
\end{figure}

\paragraf{Qubit Circuits.} 
As noted in the main text we can implement the circuits on qubits.
Each copy $\langle r, s\rangle\cong D_4$ can be realized on three qubits per edge as $|r^\rho s^\sigma\rangle\mapsto|\rho_1\rho_0\sigma\rangle$, with $\rho=2\rho_1+\rho_0$, and each $\Z_2$ factor on one qubit. This gives $3k_A-2$ qubits per edge for $G_A$ and $3k-4$ for $G_A\times G_B$. We denote the three qubits of the $j$-th copy of $D_4$ by $\rho_{j,1},\rho_{j,0},\sigma_j$. A subscript $i$ on an operator on the qubits of one copy, as in $(\bbI\otimes Z\otimes Z)_i$, means that it acts on the edge $i$, and $(\bbI\otimes Z\otimes Z)(\partial p)$ is its product over the four edges $l,t,r,b$ of the plaquette $p$. 

On the three qubits, $L^g$ and $R^g$ for $g=r,s$ are the Clifford gates of \cite{Warman:2025hov} in Fig.~\ref{fig:circuits}(a), and $L^m=R^m=X$ on the qubit of $m$. Since $A_v^{gh}=A_v^gA_v^h$, it suffices to take $g$ among the generators $r_j,s_j,m$ in the bulk, and among those in \eqref{UPQ} on the boundaries.

The plaquette stabilizers are, instead, diagonal. For two commuting operators with eigenvalues $\pm1$, let $CZ[\cdot,\cdot]$ be $-1$ if both are $-1$ and $+1$ otherwise, which for two single-qubit Pauli $Z$ gates is the usual $CZ$ gate. For each copy of $D_4$, omitting the index $j$, the plaquette stabilizers are
\be\ba \label{Sp}
    (S_p^r)^2&=(\bbI\otimes Z\otimes\bbI)(\partial p) \,, \cr
    (S_p^s)^2&=(\bbI\otimes\bbI\otimes Z)(\partial p) \,, \cr
    S_p^{r^2}&=(Z\otimes\bbI\otimes\bbI)(\partial p) \cr
    &\quad\times CZ\big[(\bbI\otimes Z\otimes Z)_l,(\bbI\otimes Z\otimes\bbI)_t\big] \cr
    &\quad\times CZ\big[(\bbI\otimes Z\otimes Z)_b,(\bbI\otimes Z\otimes\bbI)_r\big] \,,
\ea\ee
and for the $\Z_2$ factor the stabilizer is $Z_m(\partial p)$. The $(S_p^r)^2$ and $(S_p^s)^2$ plaquette stabilizers check the bits $\rho_0$ and $\sigma$ of the flux, and $S_p^{r^2}$ checks the bit $\rho_1$ (Fig.~\ref{fig:circuits}(b)), where the two $CZ$ gates account for the non-Abelian products $g_lg_t$ and $g_bg_r$.

On every edge of a boundary, the operator $B_i$ in \eqref{Ham} acts on a single edge and imposes Pauli $Z$ checks that depend on its subgroup:
\be\ba \label{edgechecks}
    \cR_A&:\ (\bbI\otimes Z\otimes Z)_j=(Z\otimes Z\otimes\bbI)_j=Z_m=1 \,, \cr
    \cS_A&:\ (Z\otimes\bbI\otimes\bbI)_j=(\bbI\otimes Z\otimes\bbI)_j=1 \,, \cr
    \cT_A&:\ (\bbI\otimes Z\otimes\bbI)_j=1 \,, \cr
    &\quad\ (Z\otimes\bbI\otimes\bbI)_j\,(\bbI\otimes\bbI\otimes Z)_{j+1}=1 \,, \cr
    &\quad\ (Z\otimes\bbI\otimes\bbI)_{k_A-1}\,Z_m=1 \,,
\ea\ee
for all copies $j$, with $j\le k_A-2$ in the fourth line, and likewise for $G_B$. Only the checks of $\cT_A$ couple different copies of $D_4$: they tie the qubit $\rho_{j,1}$ of $r_j^2$ to the qubit $\sigma_{j+1}$ of $s_{j+1}$ in the next copy, or to the qubit of $m$.

In a local basis, obtained by acting with a Hadamard gate on the qubit $\rho_{j,1}$ of each copy, the stabilizers contain only Pauli $X$, $CZ$ and $Z$ gates, which is the form of those of $D^{\omega}(\Z_2^3)$ \cite{Hu:2012wx,Yoshida:2015cia,Davydova:2025ylx}. In both bases, the stabilizers are Clifford operators.

\paragraf{Level of the Gates.} We now prove that the Clifford level of our logical gates cannot decrease by performing a change of logical basis. We will use that any two bases are related by a unitary, which preserves the ranks of operators. In general, a change of basis can change levels of the Clifford hierarchy. 

The gate $C^{k-1}Z$ lies in $\cC^{(k)}$ \cite{Gottesman:1999tea,Cui:2016bxt}, and in no lower level in any basis. \\
{\it Proof:} We use the following bound: every non-scalar $V\in\cC^{(j)}$ on $k$ qubits has 
\be \label{eq:bound}
\mathrm{rank}(V-\lambda)\ge2^{k-j}
\ee 
for every phase $\lambda$. For $j=1$ it holds since every non-scalar Pauli operator has two eigenvalues of multiplicity $2^{k-1}$. We now proceed by induction and assume \eqref{eq:bound} holds for $j$. For $V\in\cC^{(j+1)}$ and a Pauli operator $P$, the commutator $V_P=VPV^{-1}P^{-1}$ lies in $\cC^{(j)}$, and
\be \label{rank}
    V_P-1=(V-\lambda)PV^{-1}P^{-1}+\lambda P\big(V^{-1}-\lambda^{-1}\big)P^{-1} \,,
\ee
so $\mathrm{rank}(V_P-1)\le2\,\mathrm{rank}(V-\lambda)$. 
If $V_P$ is non-scalar for some $P$, the bound \eqref{eq:bound} for $V_P$ gives $\mathrm{rank}(V-\lambda)\ge2^{k-j-1}$. Otherwise $V$ commutes with every Pauli operator up to a phase, so it is a Pauli operator up to a phase, and the case $j=1$ applies. Since ranks do not depend on the basis and $\mathrm{rank}(C^{k-1}Z-1)=1<2$, the gate does not lie in $\cC^{(k-1)}$ in any basis.

\paragraf{Acknowledgments.}
We thank Dom Williamson for discussions and comments on the draft. 
This work is supported by the UKRI Frontier Research Grant, underwriting the ERC Advanced Grant ``Generalized Symmetries in Quantum Field Theory and Quantum Gravity'' (Schafer-Nameki).
We acknowledge use of Claude Code (Fable 5.2, Opus 5.5) and ChatGPT Astra-6, for improvement of arguments and writing assistance and Fable 5.2 for initial exploration of this question.

\twocolumngrid
\let\oldaddcontentsline\addcontentsline
\renewcommand{\addcontentsline}[3]{}
\bibliographystyle{ytphys}
\small
\baselineskip=.7\baselineskip
\let\bbb\bibitem\def\bibitem{\itemsep3.3pt\bbb}
\bibliography{refs}
\let\addcontentsline\oldaddcontentsline

\end{document}